%% file: template.tex
\documentclass[]{article}

\usepackage{arxiv}

\usepackage[utf8]{inputenc} 
\usepackage[T1]{fontenc}    
\usepackage{hyperref}       
\usepackage{url}            
\usepackage{booktabs}       
\usepackage{amsfonts}       
\usepackage{nicefrac}       
\usepackage{microtype}      
\usepackage{lipsum}
\usepackage{graphicx}
\usepackage{tikz}
\usepackage{tabularx} 
\usepackage{subcaption}
\usepackage{svg}            
\usepackage{layouts}        
\usepackage{comment}        
\usepackage{siunitx}

\title{A nation-wide experiment: fuel tax cuts and almost free public transport for three months in Germany - Report 4 Third wave results}

\author{
Allister Loder \\
Technical University of Munich\\
TUM School of Engineering and Design\\
Chair of Traffic Engineering and Control\\
Arcisstrasse 21, 80333 Munich \\
\texttt{allister.loder@tum.de}\\
\And
Fabienne Cantner\\
Technical University of Munich\\
TUM School of Management\\
TUMCS for Biotechnology \& Sustainability \\ 
Am Essigberg 3, 94315 Straubing\\
\texttt{fabienne.cantner@tum.de}\\
\And
Andrea Cadavid Isaza\\
Technical University of Munich\\
TUM School of Engineering and Design\\
Chair of Renewable and Sustainable Energy Systems\\
Lichtenbergstraße 4a, 85748 Garching\\
\texttt{andrea.cadavid@tum.de}\\
\And
Markus B. Siewert\\
Munich School of Politics and Public Policy\\
TUM Think Tank\\
Richard-Wagner-Straße 1, 80333 München\\
\texttt{markus.siewert@hfp.tum.de}
\And
Stefan Wurster\\
Munich School of Politics and Public Policy\\
Technical University of Munich\\
TUM School of Social Sciences and Technology\\
Professorship of Policy Analysis\\
Richard-Wagner-Straße 1, 80333 München\\
\texttt{stefan.wurster@hfp.tum.de}
\And
Sebastian Goerg\\
Technical University of Munich\\
TUM School of Management\\
TUMCS for Biotechnology \& Sustainability \\ 
Am Essigberg 3, 94315 Straubing\\
\texttt{sebastian.goerg@tum.de}\\
\And
Klaus Bogenberger\\
Technical University of Munich\\
TUM School of Engineering and Design\\
Chair of Traffic Engineering and Control\\
Arcisstrasse 21, 80333 Munich\\
\texttt{klaus.bogenberger@tum.de}\\
}

\begin{document}
\maketitle
\begin{abstract}
In spring 2022, the German federal government agreed on a set of measures that aimed at reducing households' financial burden resulting from a recent price increase, especially in energy and mobility. These measures included among others, a nation-wide public transport ticket for 9\ EUR per month and a fuel tax cut that reduced fuel prices by more than 15\,\%. In transportation policy and travel behavior research this is an almost unprecedented behavioral experiment. It allows to study not only behavioral responses in mode choice and induced demand but also to assess the effectiveness of transport policy instruments. We observe this natural experiment with a three-wave survey and an app-based travel diary on a sample 2'263 individuals; for the Munich Study, 919 participants in the survey-and-app group and 425 in the survey-only group have been successfully recruited, while 919 participants have been recruited through a professional panel provider to obtain a representative nation-wide reference group for the three-wave survey. In this fourth report we present the results of the third wave. At the end of the study, all three surveys have been completed by 1'484 participants and 642 participants completed all three surveys and used the travel diary throughout the entire study. Based on our results we conclude that when offering a 49~EUR-Ticket as a successor to the 9~EUR-Ticket and a local travel pass for 30~EUR/month more than 60~\% of all 9~EUR-Ticket owners would buy one of the two new travel passes. In other words, a substantial increase in travel pass ownership in Germany can be expected, with our modest estimate being around 20~\%. With the announcement of the introduction of a successor ticket in 2023 as well as with the prevailing high inflation, this study will continue into the year 2023 to monitor the impact on mobility and daily activities.
\end{abstract}


\section{Introduction}


In transportation research, it is quite unlikely to observe or even perform real-world experiments in terms of travel behavior or traffic flow. There are few notable exceptions: subway strikes suddenly make one important alternative mode not available anymore  \cite{Anderson2014,Adler2016}, a global pandemic changes travelers' preferences for traveling at all or traveling collectively with others \cite{Molloy2021}, or a bridge collapse forces travelers to alter their daily activities \cite{Zhu2010}. However, in 2022 the German federal government announced in response to a sharp increase in energy and consumer prices a set of measures that partially offset the cost increases for households. Among these are a public transport ticket at 9\ EUR per month\footnote{\url{https://www.bundesregierung.de/breg-de/aktuelles/9-euro-ticket-2028756}} for traveling all across Germany in public transport, except for long-distance train services (e.g., ICE, TGV, night trains), as well as a tax cut on gasoline and diesel, resulting in a cost reduction of about 15\ \% for car drivers\footnote{\url{https://www.bundesfinanzministerium.de/Content/DE/Standardartikel/Themen/Schlaglichter/Entlastungen/schnelle-spuerbare-entlastungen.html}}. Both measures were limited to three months, namely June, July and August 2022. As of end of August, more than 52\ million tickets have been sold\footnote{
\url{https://www.vdv.de/bilanz-9-euro-ticket.aspx}
}, while it seems that the fuel tax cut did not reach consumers due to generally increased fuel prices and oil companies are accused of not forwarding the tax cuts to consumers \footnote{\url{https://www.spiegel.de/wirtschaft/tankrabatt-hat-zunehmend-an-wirkung-verloren-rwi-studie-a-cb7a4e84-c943-44a3-b0d3-fcfff9ba3061?dicbo=v2-bf58f4c0d939c05bc696544c175d1063}}. 

For the Munich metropolitan region in Germany, we designed a study under the label "Mobilität.Leben" \footnote{\url{https://www.hfp.tum.de/hfp/tum-think-tank/mobilitaet-leben/}} comprising three elements: (i) a three-wave survey before, during and after the introduction of cost-saving measures; (ii) a smartphone app based measurement of travel behavior and activities during the same period; (iii) an analysis of aggregated traffic counts and mobility indicators. We will use data from 2017 as well as 2019 (pre-COVID-19) and data from shortly before the cost reduction measures for the comparison. In addition, the three-wave survey is presented to a nation-wide control group, which however does not participate in the app. The main goal of the study is to investigate the effectiveness of the cost-saving measures with focus on the behavioral impact of the 9\ EUR-ticket on mode choice \cite{ben1985discrete}, rebound effects \cite{Greening2000,Hymel2010}, and induced demand \cite{Weis2009}. Further details on the study design and the first results can be found in our first, second and third report \cite{reportone,cantner_nation-wide_2022,report_three}.

In this fourth report, we provide an update on the study participation at the end of September, the end of the initial study period of our study in Section \ref{sec:participation}. We summarize the first findings from the third wave on travel behavior before, during and after the period of the 9~EUR-Ticket and fuel tax cut in Section \ref{sec:travelbehavior}, on the willingness-to-pay for a successor ticket in Section \ref{sec:wtp}, and on the impact of price increases on households energy usage and consumer behavior.

\section{Study participation update} \label{sec:participation}

For the entire study, 2~268 participants had been successfully recruited \cite{reportone}. The entire sample comprises 1~349 participants for the \textit{Munich Study} (MS) and 919 participants from the nation-wide \textit{reference group} (RG). In the Munich Study, 425 participants participated in the survey-only group, while 919 participated in the survey-and-smartphone app group. At the end of September 2022, five participants from the Munich Study explicitly opted out. If not stated otherwise, numbers in parentheses refer to the findings from the reference group.

\begin{figure}
    \centering
    \includegraphics[width=\textwidth]{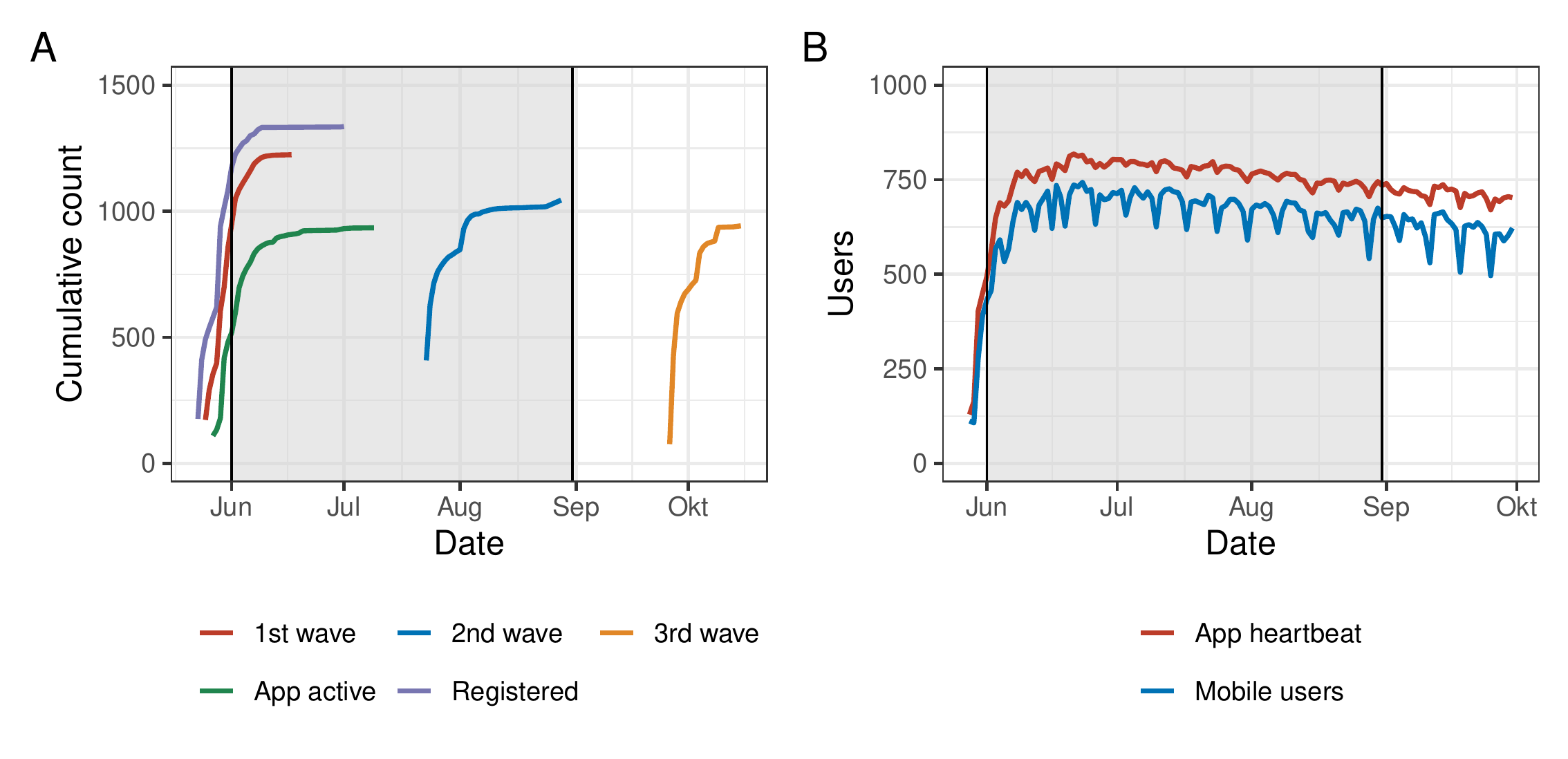}
    \caption{Study participation in October 2022 in the Munich Study. The period of the 9~EUR-Ticket and the fuel tax cut is indicated as the grey shaded area.}
    \label{fig:participation}
\end{figure}

Figure \ref{fig:participation} shows the study participation in the Munich Study before, during and after the period of the 9~EUR-Ticket and the fuel tax cut. Figure \ref{fig:participation}\textbf{A} shows that the third-wave participation, i.e., after the 9~EUR-Ticket and fuel tax cut, is around 10~\% less compared to the second wave. Compared to the number of registered participants, 70~\% completed the third wave. In the end, 727 out of 919 participants who participated in the survey-and-app group (79~\%) completed all three waves, while complete participation in the survey-only group of the Munich study was only 41~\%. The difference can be attributed to the incentive paid (30~EUR) for participation in the survey-and-app group. Nevertheless, the survey collection can be considering successful as drop outs are less than expected, arguably as many participants were intrinsically motivated to participate. In the reference group, 63~\% of the initial participants completed all three waves. 

Figure \ref{fig:participation}\textbf{B} shows the app participation during the entire study period until the end of September. It can be seen that the \textit{app heartbeat}, i.e., the app sends any signal to the server, is subsequently declining since its peak at the end of June 2022. The numbers in the app heartbeat compared to the number of activated apps are lower because not every person that successfully activated the app on their smartphone was able, e.g., for technical reasons, to initialize the tracking. The pattern of mobile users follows the expected weekly pattern with less mobility on weekends. At the end of the study, around 75~\% of all activated apps were still providing their travel diaries. Compared to a similar smartphone-based study in Switzerland \cite{axhausen_empirical_2021}, the survival rate in our study is similar when considering the planned study period. Nevertheless, the incentive paid in the Swiss study was 100 CHF for eight weeks, while we pay 30 EUR for sixteen weeks.


\section{Travel behavior and the 9~EUR-Ticket} \label{sec:travelbehavior}

\input{Figures/table_mobility.tex}

Table \ref{tab:mode_use} summarizes the mode use frequencies across the three time periods of before, during and after the 9~EUR-Ticket and fuel tax cut for all modes of public transport and the private car. Regarding public transport use, it can be seen that the shares for daily use and no use at all are relatively stable across the three periods. In other words, those how already used public transport almost daily or never did so the entire time. Contrary, the shares of 2-3 and 4-5 days per week together grow during the of the 9~EUR-Ticket and fuel tax cut by around ten percentage points in the Munich Study and around seven percentage points in the Reference Group. Regarding car use, regular use of 2 or more days per week did not vary across the three phases. Contrary, data suggests that around three to five percent of both samples changed from less than once a week to once a week from the before to the during period. One hypothesis could be the holiday season or the fuel tax cut and the subsequent lower prices.

Some participants increased public transport usage compared to the \textit{before} period despite the discontinuation of the 9~EUR-Ticket: 8.91~\% (RG: 5.83~\%) reported that they were using public transport more frequent in the \textit{after} period compared to the \textit{before} period. Corroborating this finding using the mode use frequencies shows that 24.7~\% (RG: 17.66~\%) increased public transport usage. Regarding travel pass ownership (at least a weekly pass), 7.28~\% (RG: 4.51~\%) of previous non-owners subscribed to a travel pass in September. 53.00~\% (RG: 32.93~\%) of the respondents stated that they replaced some car trips with public transport; these respondents further  stated that in the \textit{after} period 15.38~\% (RG: 27.67~\%) of weekly car trips are still being substituted by public transport, which is down from 34.75~\% (RG: 45.09~\%) in the during phase. Nevertheless, 24.03~\% (RG: 15.27~\%) reduced public transport usage in the \textit{after} period compared to the \textit{before} period. Of those 67.06~\% (RG: 78.90~\%) who did not change their public transport use when comparing the \textit{before} and \textit{after} phase, 29.72~\% (RG: 12.61~\%) used public transport more in the \textit{during} period. Generally, 44.67~\% (RG: 22.98~\%) indicated that they were using public transport less in the \textit{after} period compared to the \textit{during} period.

The discontinuation of the 9~EUR-Ticket and fuel tax cut changed the lives of many: 38.39~\% (RG: 28.69~\%) of respondents argued that after this discontinuation parts of their daily routines have been changed. Here, in particular, 48.47~\% (RG: 29.33~\%) stated that they were more mobile during the 9~EUR-Ticket and fuel tax cut period; of all more mobile respondents 80.37~\% (RG: 73.63~\%) named the 9~EUR-Ticket as one of the key reasons for their increased mobility, while participation in more social events was named by 46.26~\% (RG: 41.52~\%), note that multiple answers were possible.

When summarizing the above findings, it can be concluded that travel behavior in terms of public transport use changed as expected with an increase in the during phase. Interestingly, we find evidence in the survey data for a small hysteresis effect, i.e. some maintained (parts) their increased public transport from the during period, where its precise quantification depends on the chosen measure.

\section{Willingness-to-pay for a successor ticket} \label{sec:wtp}

\begin{figure}
    \centering
    \includegraphics[width=\textwidth]{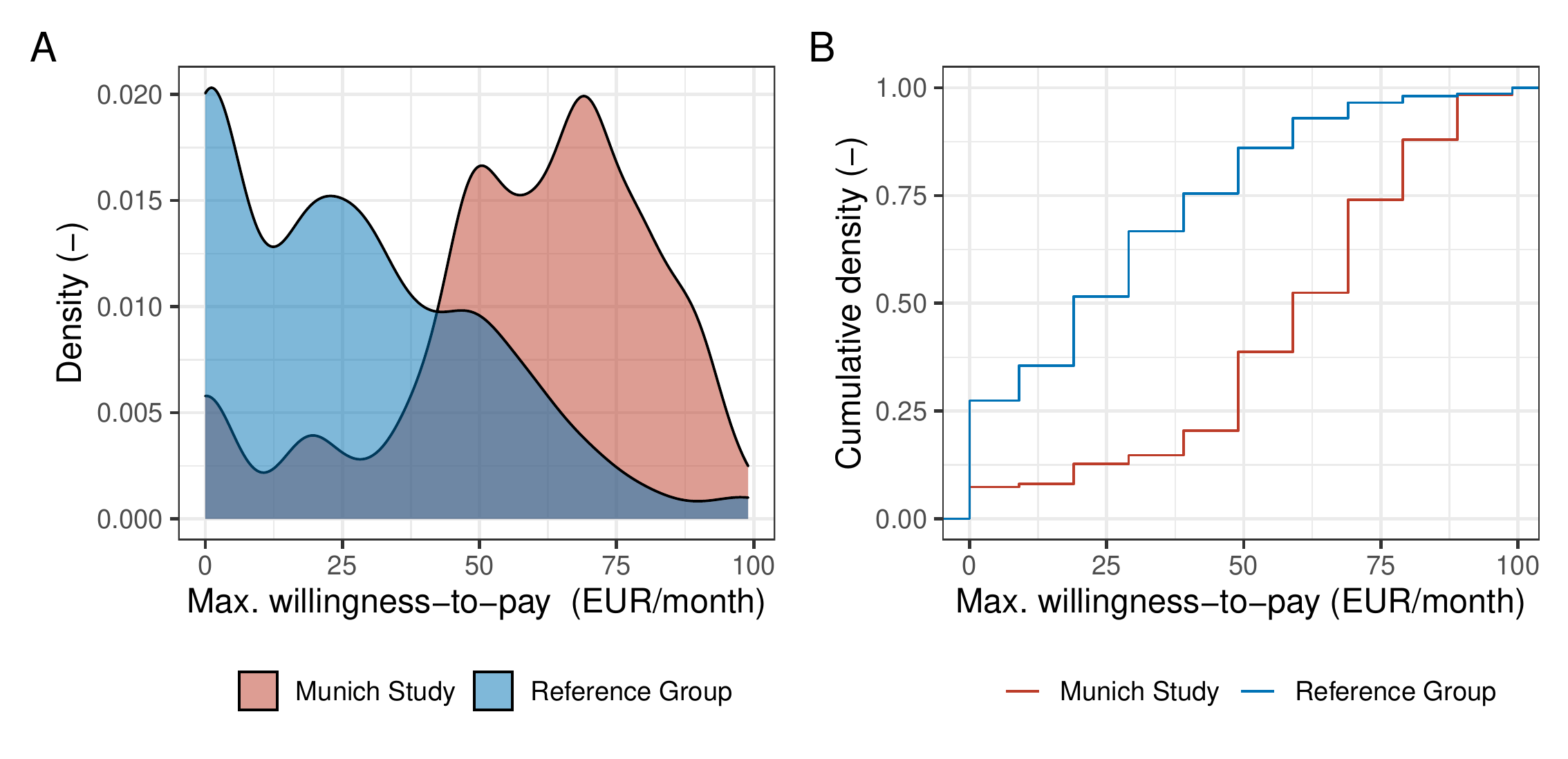}
    \caption{Distributions of the willingness-to-pay for a successor ticket to the 9~EUR-Ticket in the Munich Study and the Reference Group. \textbf{A} shows the density and \textbf{B} the cumulative distributions.}
    \label{fig:wtp_direct}
\end{figure}

Participants in the third wave have been asked in a twofold way to state their consumer preferences for a successor to the 9~EUR-ticket. In the first approach, participants were asked to state their maximum willingness-to-pay for a successor to the 9~EUR-Ticket. The resulting distributions are shown in Figure \ref{fig:wtp_direct}. A clear difference between the Munich Study and the Reference Group can be seen for both, the kernel density estimate in Figure \ref{fig:wtp_direct}\textbf{A} as well as the cumulative density plot in Figure \ref{fig:wtp_direct}\textbf{B}. The Munich Study has on average a much larger maximum willingness-to-pay compared to the Reference Group. Based on our findings in  our second-wave report \cite{report_three}, this can be attributed to the higher income levels in our Munich Study as well as due to the more public transport oriented and experienced sample. Generally, we can conclude that a successor ticket priced at 49~EUR would be acceptable for around 60~\% (RG: 15~\%) and a successor ticket priced at 69~EUR only for 25~\% (RG: 5~\%). Note that the 9~EUR-Ticket has been bought by 83~\% (RG: 45~\%) of participants. However, such maximum willingess-to-pay must not directly translate into purchasing decisions. 

In order to understand the purchasing decisions of consumers in 2023, we performed a stated preference experiment in the third wave. In this experiments, respondents were tasked to decide which of the following options they would buy:

\begin{itemize}
    \item No travel pass
    \item A local or regional travel pass, covering the area of existing transit districts, that is priced at 19 or 29~EUR per month, following the idea of providing mobility for 1~EUR/day.
    \item The Deutschlandabo as the successor to the 9~EUR-Ticket that is priced at 49, 59, 69 or 99~EUR per month, following the currently discussed price points for that ticket. 
    \item The Deutschlandabo including long-distance services, i.e., valid for \textbf{all} public transport services in Germany, priced at 249 or 349~EUR per month. This ticket is similar to the BahnCard~100, but includes all local and regional public transport services as well.
    \item A distance fare system with a price cap at the price for the Deutschlandabo including long-distance services. The distance fares are 10 or 20~EUR per 100~km and is aligned with prices typically obtained from using the half-fare card BahnCard~50.
\end{itemize}

We employed a full-factorial design, in total 32 choice sets, and grouped them into four blocks, i.e., each respondent is tasked with in total of eight decision scenarios, where only price attributes are varying. Using the data from the Reference Group to obtain as nation-wide representative parameter estimates as possible for the consumer decisions, we estimated a mixed logit model \cite{Train2009} using the R package mixl \cite{molloy_mixl_2019}. Table \ref{tab:model_estimates} summarizes the parameter estimates.

The parameter estimates in Table \ref{tab:model_estimates} are first results and further model development is required in future research. However, these first results allow to understand already the basic mechanism in the decision making for travel pass ownership. Generally, all parameter estimates for the prices are negative and significant as expected. Interestingly, we find a negative effect of being male on local travel pass and a positive effect of being male on the Deutschlandabo. A possible explanation could be that men might commute longer distances, eventually beyond the borders of the existing transit districts. Living in an urban environment (following the RegioStar classification) increases the likelihood of buying any travel pass. Perhaps surprisingly, we find no significant income effect in the decision for the Deutschlandabo, while it exists in the choices for the Deutschlandabo incl. long-distance services and the distance fare ticket. In addition, we find that car ownership impacts negatively the choices of local pass and Deutschlandabo, while a positive effect can be seen for the choice of Deutschlandabo incl. long-distance services. Together with the significant income effect, the latter could be an indication of wealthy households opting for several mobility tools in parallel. 

\begin{table}[]
\caption{Parameter estimates for the travel pass ownership model. Note that only the observations from the Reference Group are used to obtained nation-wide representative parameter estimates as possible. The alternative no travel pass is set at the reference point.}
\label{tab:model_estimates}
\scriptsize
\begin{tabularx}{\textwidth}{X | c | c | c | c | c}
\toprule
Parameter & No pass (base) & Local pass & Deutschlandabo & Deutschlandabo & Distance fare \\
& & & & incl. & \\
& & & & long-distance services & \\
\midrule
ASC & & 2.95 (3.68) &1.07 (1.27) & -8.06 (-5.38) & -5.76 (-4.89) \\
$\sigma_{ASC}$  & & 4.47 (13.56) & -3.97 (-10.70) & 4.90 (6.53) & -4.89 (-10.17) \\
Price& & -0.14 (-7.43) & -0.06 (-7.10) & -0.01 (-2.38) & -0.1 (-3.65) \\
Male & &- 0.78 (-2.19) & 1.22 (2.98) & 0.61 (0.85) & -0.14 (-0.25) \\
Urban & & 2.42 (5.88) & 1.24 (3.01) & 1.74 (1.58) & 0.99 (1.63) \\
Car ownership & & -3.566 (6.21) & -0.99 (-2.20) & 1.5 (1.57) & -0.36 (-0.55)\\
Income less than 1499~EUR per month & & base & base & base & base \\
Income from 1500 to 2499~EUR per month  & & 0.29 (0.50) & 0.93 (1.50) & 0.99 (0.81) & 1.65 (2.53) \\
Income from 2500 to 3999~EUR per month  & & 1.25 (1.90) & 0.09 (0.15) & 2.16 (2.10) & 2.96 (3.56)\\
Income more than 4000~EUR per month  & & 0.82 (1.4) & 0.08 (0.12) & 2.24 (1.92) & 2.58 (3.57) \\
\midrule
\multicolumn{6}{l}{Number of decision makers: 581}\\
\multicolumn{6}{l}{Number of choices: 4598}\\
\multicolumn{6}{l}{{$\mathcal{L}\mathcal{L}\left(0\right)=7400.2$}}\\
\multicolumn{6}{l}{{$\mathcal{L}\mathcal{L}\left(final\right)=3128.6$}}\\
\multicolumn{6}{l}{Adj. $\rho^2 = 0.58 $ }\\
 \bottomrule
\end{tabularx}
\end{table}

\begin{figure}
    \centering
    \includegraphics[width=\textwidth]{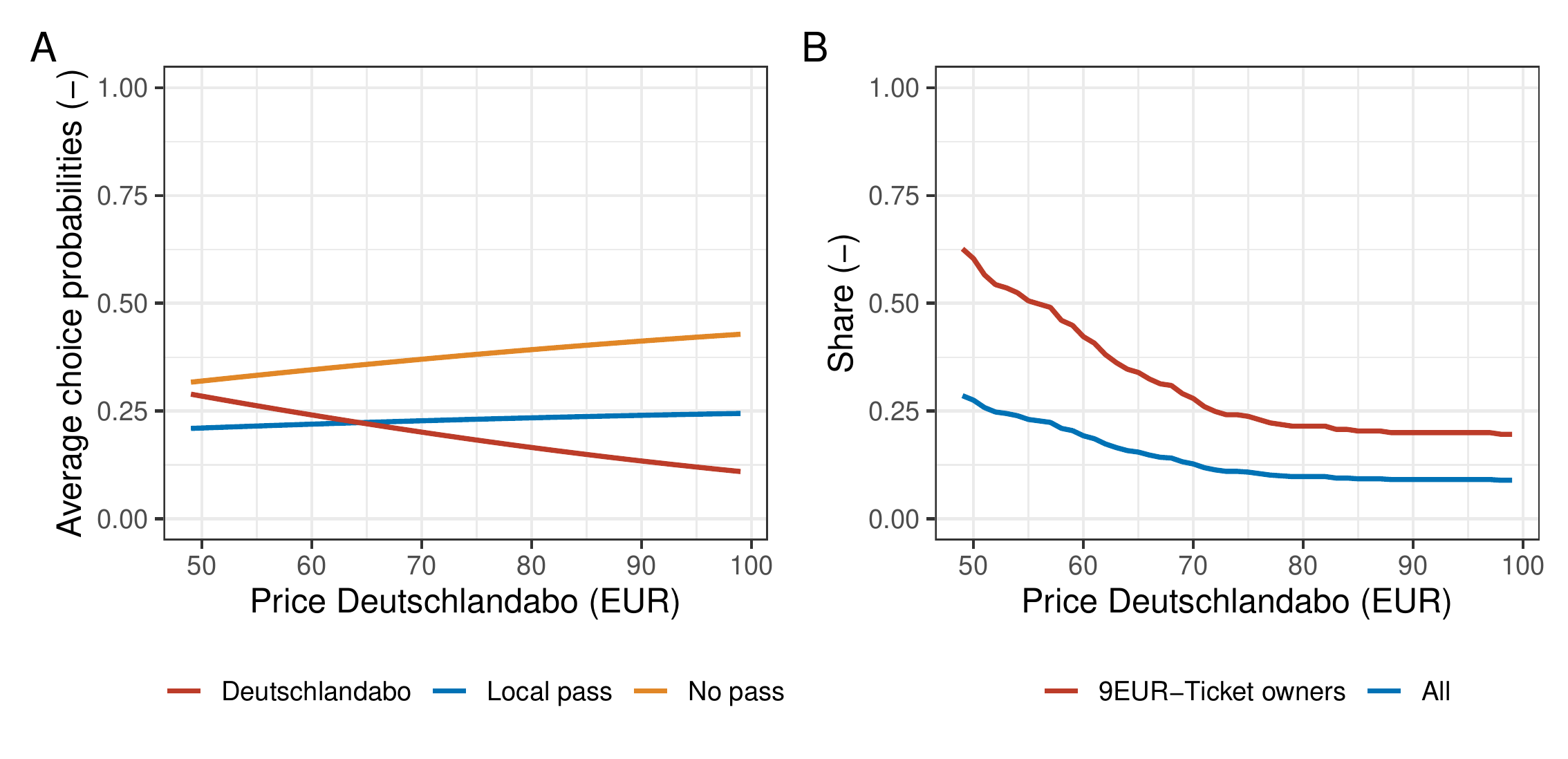}
    \caption{Estimated choices for the ownership of new travel passes in Germany. \textbf{A} the average choice probabilities and \textbf{B} share of customers who buy one of the new tickets of all previous 9~EUR-Ticket owners.}
    \label{fig:wtp_mixl}
\end{figure}

We use the parameter estimates from Table \ref{tab:model_estimates} to simulate the choice outcomes as a function of the price of the Deutschlandabo. The results are shown in Figure \ref{fig:wtp_mixl}. Figure \ref{fig:wtp_mixl}\textbf{A} shows the average choice probabilities for no travel pass, the local travel pass (at 30~EUR/month) and the Deutschlandabo. It can be seen that on average, at no currently discussed price level the average choice probability surpasses the probabilities for obtaining no travel pass at all, i.e., at no price level can it be expected that more than half of the population will buy a travel pass. Interestingly, we see that between 60 and 70~EUR/month the average choice probability for the Deutschlandabo falls below the local travel pass, i.e., it is becoming less attractive for the average citizen. Figure \ref{fig:wtp_mixl} shows the shares of ownership among all and among previous 9~EUR-Ticket owners. When priced at 49~EUR/month for the 9~EUR-Ticket successor and 30~EUR/month for a local travel pass, more than 60~\% of previous 9~EUR-Ticket owners would purchase one of the two new tickets.  At this price level, our sample would see an increase in travel pass ownership by more than 20~\%, while at higher price levels ownership levels would drop below current ownership levels. Note that the average monthly ticket currently costs around 55~EUR in Germany, making a 100~EUR ticket very unattractive to many.

\section{Finances and energy}\label{sec:household}

Effects of the persisting price increases in Germany in 2022 can be seen in the survey data. In May 2022, 57.81~\% (RG: 45.31~\%) of households were able to make savings of their income; of these, 87.88~\% (RG: 77.39~\%) can still do in September, leaving 10.10~\% (RG: 16.48~\%) making no savings anymore and 2.02~\% (RG: 6.13~\%) drawing money from savings. Using a five-point Likert-scale, respondents were asked to agree with the statement that due to the increased prices, they have to do without many other things in life. In May 2022, 16~\% (RG: 41~\%) of respondents agreed with the statement, but in September 2022, 26~\% (RG: 50~\%) agreed, i.e., by selecting the two highest options on the scale. Generally, 41~\% (RG: 31~\%) of respondents agreed stronger with the statement in September compared to May, i.e., selected a higher option on the scale in September than in June.  

Regarding the energy crisis, 43~\% (51~\%) of respondents are concerned about the energy security in the upcoming winter. Here, 75~\% (65~\%) of respondents stated that they already took measures to reduce energy costs, where 90~\% (83~\%) named reducing electricity and heating energy consumption as their primary action. From the participants who had already received their utility bills from last winter, 49~\% (48~\%) of them did receive a back payment, which was mainly under 250 Euros  55~\% (51~\%) or between 250 and 500 Euros 33~\% (29~\%). In contrast, 32~\% (39~\%) of the respondents had to pay extra, again mostly less than 250 euros 68~\% (64~\%) followed by repayments between 250 and 500 euros 19~\% (16~\%). When asked about their current energy running costs, 51~\% (49~\%) of the participants indicated that the monthly prepayment for electricity had not changed lately and for heating 51~\% (57~\%) also saw no change. On the contrary, 36~\% (42~\%)  of the respondents did get a change on their monthly prepayment for electricity and 29~\% (27~\%) for their heating costs.

\section{Discussion and outlook}

In this fourth report, we have provided some first insights into the third wave of our study "Mobilität.Leben", which marks the end of the first phase of our study. Given the ongoing energy crisis and the likely introduction of a successor to the 9~EUR-Ticket, we will bring this study into its second phase and continue until the second quarter of the year 2023. The first findings of this report can summarized as follows

\begin{itemize}
    \item We completed the first phase of our study successfully with around 1000 (RG: 581) completed three survey waves and more than 600 participants who completed the three waves and participated in the smartphone app. 
    \item  Since the end of the 9~EUR-Ticket and fuel tax cut, a hysteresis effect seems to exist in terms of public transport usage for some respondents, e.g., reflected a more frequent use of public transport of 8.9~\% (RG: 5.8~\%) after the 9~EUR-Ticket period compared to the time before as well as 7.3~\% (RG: 4.5~\%) of all respondents subscribing for the first time to a travel pass. Nevertheless, the changes in travel behavior during the 9~EUR-Ticket period are not that substantial, which is in line with experience from other studies exploring free-fare public transport \cite{keblowski_why_2020}.
    \item The currently discussed price for the successor to the 9~EUR-Ticket of around 49~EUR/month would see a decrease in purchases compared to the 9~EUR-Ticket, but based on our findings it can be expected that more than 60~\% of all previous 9~EUR-Ticket owners would either buy the successor ticket for 49~EUR-Ticket or a new local travel pass at around 30~EUR/month, which would lead to an increase of travel pass ownership in Germany of more than 20~\%. 
    \item Interestingly, we do find an income effect in the purchase decision for a successor to the 9~EUR-Ticket, while other factors like living in an urban environment, having no car or being male increase the likelihood for the ticket purchase. Nevertheless, we reported an income effect for the willingness-to-pay in our report on the second wave \cite{report_three}, where we see that lower incomes prefer a 9~EUR-Ticket successor priced at around 30~EUR/month, but this effect vanishes for middle and higher incomes. As this price level has not been used in the stated-preference experiment lower income individuals would arguably not choose the successor ticket, but either the local travel pass or no pass instead.
\end{itemize}

In closing, it should be noted that this report does not present the final results of our study, but rather a snapshot of our current research as we are constantly analyzing the data further. Therefore, the presented findings should be considered as preliminary and indicative. 

\section*{Acknowledgements}

The authors would like to thank the TUM Think Tank at the Munich School of Politics and Public Policy led by Urs Gasser for their financial and organizational support and the TUM Board of Management for supporting personally the genesis of the project. The authors thank the company MOTIONTAG for their efforts in producing the app at unprecedented speed. Further, the authors would like thank everyone who supported us in recruiting participants, especially Oliver May-Beckmann (M Cube) and Ulrich Meyer (TUM), respectively.

\bibliographystyle{unsrt}  
\bibliography{references}  



\end{document}

%% file: Figures/table_mobility.tex
\begin{table}[]
\caption{Mode usage frequencies per week for car and public transport before, during and after the period of the 9~EUR-Ticket and the fuel tax cut.}
\label{tab:mode_use}
\scriptsize
\begin{tabularx}{\textwidth}{Xrrrrrrcrrrrrr}
\toprule
 & \multicolumn{6}{c}{Munich Study (N=901)} & \multicolumn{1}{c}{} & \multicolumn{6}{c}{Reference Group (N=583)} \\ \cline{2-7} \cline{9-14} 
 & \multicolumn{2}{c}{Before} & \multicolumn{2}{c}{During} & \multicolumn{2}{c}{After} & \multicolumn{1}{c}{} & \multicolumn{2}{c}{Before} & \multicolumn{2}{c}{During} & \multicolumn{2}{c}{After} \\ \cline{2-7} \cline{9-14} 
 & \multicolumn{1}{c}{Car} & \multicolumn{1}{c}{PT} & \multicolumn{1}{c}{Car} & \multicolumn{1}{c}{PT} & \multicolumn{1}{c}{Car} & \multicolumn{1}{c}{PT} & \multicolumn{1}{c}{} & \multicolumn{1}{c}{Car} & \multicolumn{1}{c}{PT} & \multicolumn{1}{c}{Car} & \multicolumn{1}{c}{PT} & \multicolumn{1}{c}{Car} & \multicolumn{1}{c}{PT} \\ \midrule 
Daily & 6.01 & 10.79 & 5.80 & 11.05 & 6.33 & 8.21 &  & 34.27 & 7.90 & 31.90 & 7.20 & 29.50 & 7.20 \\
4-5 days/week & 11.56 & 16.24 & 10.94 & 22.10 & 11.84 & 16.76 &  & 19.56 & 7.56 & 20.41 & 10.12 & 19.04 & 7.20 \\
2-3 days/week & 19.50 & 22.14 & 20.09 & 28.57 & 20.98 & 24.03 &  & 19.35 & 9.79 & 19.04 & 13.21 & 22.81 & 9.26 \\
Once a week & 13.95 & 15.13 & 18.30 & 16.96 & 18.29 & 19.23 &  & 6.85 & 6.19 & 10.98 & 9.26 & 9.61 & 10.12 \\
Less than once a week & 26.87 & 28.48 & 23.44 & 14.06 & 23.92 & 20.16 &  & 8.47 & 30.07 & 5.32 & 17.84 & 6.00 & 18.70 \\
Never & 22.11 & 7.23 & 21.43 & 7.25 & 18.64 & 11.61 &  & 11.49 & 38.49 & 12.35 & 42.37 & 13.04 & 47.51 \\ \bottomrule
\end{tabularx}
\end{table}